\documentclass[10pt]{iopart}

\usepackage{siunitx}
\usepackage{amsfonts}
\usepackage{amssymb}
\usepackage{graphicx}
\usepackage[textsize=scriptsize, textwidth=50]{todonotes}
\usepackage[autostyle]{csquotes} 
\usepackage{bm}
\usepackage{subcaption} 
\usepackage{booktabs}
\usepackage{colortbl} 
\usepackage{makecell} 
\usepackage{import} 

\newcommand{\soutSR}[1]{}

\usepackage[switch]{lineno} 
\DeclareCaptionFormat{custom}
{%
    \textbf{#1#2}\textit{#3}
}

\usepackage[
    colorlinks=true,
    linkcolor=blue,
    citecolor=blue,
    urlcolor=blue,
    pdfauthor={Sascha Rienaecker},
    bookmarksopen=true,
    ]{hyperref}
    
\usepackage[backend=biber,
    bibstyle=numeric-comp,
    citestyle=numeric-comp, %
    maxcitenames=1,
    sorting=none, 
    giveninits=true,
    ]{biblatex}

\usepackage{biblatex-shortfields}
\usepackage{datetime}
\newdateformat{monthyear}{\monthname[\THEMONTH] \THEYEAR}

\AtEveryBibitem{%
\clearfield{url}%
\clearfield{urlyear}%
\clearfield{title}%
}

\newcommand{\low}[1]{_\mathrm{#1}}

\newcommand{\up}[1]{\mathrm{#1}}

\newcommand{\refig}[1]{Fig.\,\ref{fig:#1}}
\newcommand{\refsec}[1]{Sec.\,\ref{sec:#1}}

\newcommand{\reftab}[1]{Tab.\,\ref{tab:#1}}

\newcommand{\colpslashm}{\textcolor{red}{+}/\textcolor{blue}{-}} 

\newcommand{\ErxB}{E_r \times B}
\newcommand{\BxgradB}{B \! \times \! \nabla B}

\newcommand{\PLH}{P\low{L \rightarrow H}}

\def\figpath{./figures}

\graphicspath{
              {\figpath}
              }

\usepackage[final, authormarkup=none=color, commandnameprefix=ifneeded]{changes} 
\definechangesauthor[name={SR}, color=red]{SR}

\makeatletter
\newcommand{\mainmatter}{%
\setcounter{footnote}{0}%
\patchcmd{\@makefntext}{\fnsymbol}{\alph}{}{}%
\patchcmd{\@thefnmark}{\fnsymbol}{\alph}{}{}%
\def\@makefnmark{\textsuperscript{\alph{footnote}}}%
}
\makeatother

\begin{document}

\title{Impact of magnetic drift configuration on the edge radial electric field in the TCV tokamak}


\author{S. Rienäcker$^1$, L. Vermare$^1$, P. Hennequin$^1$, C. Honoré$^1$, B. Labit$^2$
        S. Coda$^2$, L. Frassinetti$^3$, B. Vincent$^2$, O. Panico$^2$, Y. Wang$^2$, K. Singh$^2$, the TCV team\footnote{See author list of C. Theiler et al 2026 Nucl. Fusion \href{https://doi.org/10.1088/1741-4326/ae6d12}{66 116007}}
        and the EUROfusion Tokamak Exploitation team\footnote{See author list of N. Vianello et al 2026 Nucl. Fusion \href{https://doi.org/10.1088/1741-4326/ae71ec}{66 116010}}
        }

\address{$^1$ Laboratoire de Physique des Plasmas (LPP), CNRS, Sorbonne Université, École polytechnique, Institut Polytechnique de Paris, Palaiseau, France}
\address{$^2$ Ecole Polytechnique Fédérale de Lausanne, Swiss Plasma Center, Lausanne, Switzerland}
\address{$^3$ Division of Fusion Plasma Physics, KTH Royal Institute of Technology, Stockholm, Swede}

\ead{sascha.rienacker@lpp.polytechnique.fr}
\vspace{10pt}
\begin{indented}
\item[]\monthyear\today 
\end{indented}


\begin{abstract}
The edge radial electric ($E_r$) in the \textit{Tokamak à Configuration Variable} (TCV) is compared in matched L-mode discharges with opposite ion magnetic drift directions (favorable versus unfavorable $\BxgradB$ configurations). As previously reported on the WEST and AUG tokamaks, the $E_r$ profile---measured by Doppler backscattering (DBS)---exhibits a \enquote{well} just inside the separatrix in the favorable drift case, which is absent or less pronounced in the unfavorable counterpart. This observation holds over a broad range of plasma conditions, notably also in Ohmic discharges with nearly identical edge density and temperature profiles.     
Density fluctuation characteristics inferred from DBS are not drastically different: Under Ohmic heating, edge fluctuation levels tend to be higher in the favorable configuration, while radial correlation lengths are similar. 
The edge $E_r$ difference appears uncorrelated with carbon toroidal rotation behavior. 
Reducing density, or increasing auxiliary heating power tends to accentuate the edge $E_r$ disparity. Plasma current has little impact on both favorable and unfavorable $E_r$ profiles---in contrast to results from WEST.
Approaching the L-H transition via auxiliary heating, edge $\ErxB$ shear and pressure grow more readily in the favorable configuration, and DBS fluctuation levels are reduced relative to the unfavorable case.
Overall, our results confirm that the edge $E_r$ sensitivity to magnetic drift configuration is a robust, multi-machine phenomenon, with a plausible connection to confinement level and H-mode access.
\end{abstract}

%
%
\submitto{\PPCF}
%
%
\ioptwocol

\mainmatter

\section{Introduction}\label{sec:intro}

The radial electric field ($E_r$) plays a fundamental role in regulating turbulent transport in tokamak plasmas via $\ErxB$ flow shear\,\cite{Biglari_1990, Burrell_1997, Terry_2000}. In the plasma edge region, which is critical for overall confinement, identifying the dominant mechanisms shaping the $E_r$ structure—particularly the so-called $E_r$ \enquote{well}—remains challenging. Similarly, a comprehensive understanding of how edge $\ErxB$ shear relates to high confinement access, such as the low-to-high confinement mode (L-H) transition, is still lacking. 
Addressing these fundamental aspects is essential for improving our understanding of transport and confinement regime transitions in fusion plasmas.

An outstanding knowledge gap in this context is the universal asymmetry with regard to H-mode access observed across many tokamaks\,\cite{Wagner_1985, Carlstrom_1994, Hubbard_1998, Ryter_1995, Chen_2018}: The L-H transition power threshold ($\PLH$) increases by a factor of approximately 2--3 when the ion magnetic drift, or $\BxgradB$ direction, points away from, rather than toward, the active X-point. These configurations are termed unfavorable (UNFAV) and favorable (FAV), respectively.
Besides $\PLH$, the edge $E_r$ profile also exhibits a marked sensitivity to drift configuration. As documented on the WEST and AUG tokamaks, the L-mode $E_r$ well and associated $\ErxB$ shear are typically more pronounced in plasmas with FAV drift than in their UNFAV counterparts\,\cite{Schirmer_2006, Vermare_2022, Plank_2023}. A connection between H-mode access and $E_r$ asymmetries appears plausible: Stronger edge flow shear in FAV might promote turbulence suppression, facilitating the onset of the L-H transition and reducing $\PLH$ compared to the UNFAV case.
However, such a causal link has not yet been established, and the primary cause of the FAV/UNFAV asymmetries remains unclear.
Candidate mechanisms for explaining the sensitivities of $\PLH$ and/or $E_r$ to the drift configuration include: edge or scrape-off layer (SOL) flows driven by cross-field transport\,\cite{LaBombard_2004,LaBombard_2005,LaBombard_2008,Aydemir_2009}, edge ion-orbit losses\,\cite{Chankin_1993, Shaing_2002, Brzozowski_2019}, or Reynolds stress associated with magnetic-shear-induced eddy tilting\,\cite{Fedorczak_2012,Fedorczak_2013, Manz_2018, Peret_2022, Grover_2024}.
Recent gyrokinetic simulations in realistic AUG geometry using the \texttt{GENE-X} code\,\cite{Frei_2026} reproduce the experimental $E_r$ difference and support the latter, turbulence-driven explanation. 
Meanwhile, recent 2D mean-field simulations carried out using the \texttt{SOLEDGE3X} code, both under COMPASS\,\cite{Lambresa_inprep} and WEST\,\cite{ElSaifi_inprep} plasma conditions, reveal $E_r$ asymmetries consistent with the experimental trend. These results were obtained in the absence of self-consistent fluctuation dynamics or ion orbit losses. Further turbulent simulations with \texttt{FELTOR}\,\cite{Gerru_2022_phdthesis} and \texttt{GBS}\,\cite{DeLucca_2026_preprint} qualitatively capture the $E_r$ difference, although, below a certain heating power, the \texttt{GBS} simulations exhibit no or the opposite trend\,\cite{DeLucca_2026_preprint,Krutkin_2026_inprep}. 
Despite these recent studies, a generally accepted understanding of the $E_r$ difference between FAV and UNFAV is still lacking.

Detailed experimental comparisons between FAV and UNFAV configurations can pinpoint key parameters governing the drift configuration impact on both $E_r$ and transport. Such studies provide a foundation for validating and informing modeling efforts to elucidate the underlying mechanism and its possible connection to the H-mode access asymmetry.
To this end, we study and compare the edge $E_r$ and other plasma parameters in matched FAV and UNFAV discharges in the \textit{Tokamak à Configuration Variable} (TCV).
$E_r$ measurements are performed using Doppler backscattering, complementing similar studies previously conducted on WEST\,\cite{Vermare_2022} and AUG\,\cite{Schirmer_2006,Plank_2023}.

The paper is organized as follows: \refsec{method} lays out the experimental method. \refsec{tcv_fav_unfav} presents an in-depth, reference FAV/UNFAV comparison. \refsec{ip_scan}, \refsec{density_scan} and \refsec{heating_scan} examine the role of plasma current, density and auxiliary heating, respectively. Results are discussed in \refsec{discussion}, and summarized in \refsec{summary}.

\section{Experimental method}\label{sec:method}

The TCV tokamak\,\cite{Theiler_2026} (major radius $R = \SI{0.88}{m}$, 
minor radius $a = \SI{0.25}{m}$, on-axis magnetic field $|B_0| < \SI{1.54}{T}$, plasma current $|I_p| < \SI{1}{MA}$) is well suited for the present study given its highly flexible plasma geometry, with arbitrary signs of $B_0$ and $I_p$. Compared to past similar studies on the full-tungsten devices WEST and AUG, TCV offers complementary characteristics as a smaller, carbon-wall tokamak.

As on WEST and AUG, the turbulence perpendicular velocity $v_\perp$---a proxy for $E_r$---is measured in TCV via Doppler backscattering (DBS)\,\cite{Hirsch_2001, Conway_2004, Hennequin_2006}. This active diagnostic technique uses a microwave beam, tilted with respect to the cutoff surface normal, to detect the backscatter from density fluctuations. 
Power and Doppler shift of the detected complex signal reflect the fluctuations' intensity and perpendicular velocity ($v_\perp$), respectively. The scattering structures are radially localized near the beam turning point, with a perpendicular wavenumber ($k_\perp$) selected by the scattering geometry (\enquote{Bragg rule}). Beam turning point and $k_\perp$ are estimated via beamtracing\,\cite{Honore_2006}.
Neglecting a possible turbulence-intrinsic (\enquote{phase}) velocity compared to the background $\ErxB$ flow, $v_\perp \approx E_r / B$. Different radial locations are sampled by stepping the probing microwave frequency.

TCV is equipped with a dual-channel V-band DBS system\,\cite{Rienaecker_2025}, enabling radial profile measurements of $v_\perp$ from the outer core to the edge region. Under stationary plasma conditions, a 40-point $v_\perp$ profile is typically acquired every \SI{100}{ms}.
Moreover, combining both DBS channels enables spatial correlation analysis\,\cite{Panico_2026_inprep}, discussed in \refsec{radcorr}. The DBS beam is launched downward from an upper port on the low-field side, granting access to the first quadrant of the plasma cross-section. 
Further details of the DBS system and data processing are provided in Ref.\,\cite{Rienaecker_2025}.
Radial profiles are shown as a function of the flux surface label ($\rho_\psi$), defined as the square root of the normalized poloidal flux. The sign convention for $v_\perp$ is positive (negative) when directed along the ion (electron) diamagnetic direction (i.e., $E_r$ and $v_\perp$ have the same sign). Otherwise, the standard COCOS17\,\cite{Sauter_2013} toroidal coordinate conventions apply, where positive $I_p$ or $B_0$ is anti-clockwise when viewed from above.
Electron density ($n_e$) and temperature ($T_e$) measurements are performed via Thomson scattering (TS)\,\cite{Blanchard_2019}. 
The ion temperature ($T_i$) and toroidal rotation ($V_\varphi$) of the dominant impurity species (carbon, C$^{6+}$) are obtained by charge exchange recombination spectroscopy (CXRS)\,\cite{Bagnato_2023}. The CX reactions are triggered non-intrusively using TCV's low-power, low-torque diagnostic neutral beam (DNBI)\,\cite{Karpushov_2009}.

\section{Characterization of reference discharges}\label{sec:tcv_fav_unfav}

\subsection{Discharge conditions}

\begin{figure}[htbp]
    \centering
    \begin{minipage}[c]{\linewidth} 
        \centering
        \includegraphics[height=7cm]{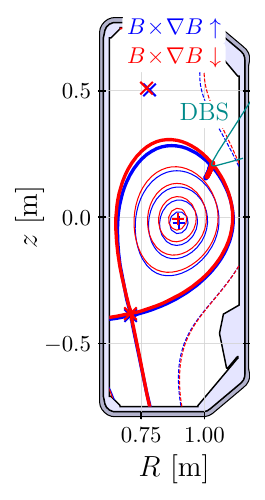}
        \caption{$\rho_\psi$ contours of the reference equilibria. DBS measurement locations are indicated together with an example beam trajectory in the first quadrant.}\label{fig:eq_fav_unfav_reference}
    \end{minipage}%
    \hfill
    \begin{minipage}[c]{\linewidth} 
        \centering
		{\small
		\setlength{\tabcolsep}{3pt} 
        \begin{tabular}{l| >{\color{blue}}c| >{\color{red}}c}
            \toprule
            Drift & UNFAV & FAV \\
            \midrule
            $\#$                    &    86594  &  86593   \\
            $I_p$\,[kA]             &     $+\textcolor{black}{210}$    &   $-\textcolor{black}{210}$     \\ 
            $B_0$\,[T]              &     $+\textcolor{black}{1.44}$       &  $-\textcolor{black}{1.44}$ \\ 
            $q_{95}$                &       $\textcolor{black}{3.6}$ & $\textcolor{black}{3.6}$  \\
			$\bar{n}_{l}$ [10$^{19}$m$^{-3}$] & 	$\textcolor{black}{3.6}$ & $\textcolor{black}{3.6}$  \\
			$\bar{n}_l / n_G$			&   $\textcolor{black}{0.29}$ & $\textcolor{black}{0.29}$  \\
			$W\low{tot}$ [kJ]			&	6.5							&	7 \\
			$\beta_N$					&	1.04						& 1.06 \\
			gas [10$^{21}$ p./s]		&	0.17						& 0.27 \\


            \bottomrule
        \end{tabular}
		}
        \captionof{table}{Main discharge parameters.}\label{tab:tcv_lsn_fav_unfav}
        \end{minipage}
\end{figure}
A representative pair of FAV/UNFAV Ohmic deuterium discharges is compared in detail. They have a lower single-null (LSN) geometry (\refig{eq_fav_unfav_reference}) with opposite $\BxgradB$ and $I_p$ directions, while their shape and other macroscopic parameters are largely matched. The primary plasma parameters (listed in \reftab{tcv_lsn_fav_unfav}) correspond to intermediate Ohmic operating conditions on TCV.
While stored energy ($W\low{tot}$) and normalized beta $\beta_N$ values are similar, the injected gas rate is higher by a factor of 1.6 in FAV. Turbulence wavenumbers sampled by the DBS at the very edge ($\rho_\psi = 1.0$--$0.95$) lie within $k_\perp \approx 6$--$\SI{8}{\per \centi \m}$, or $k_\perp \rho_s \approx 0.5$--$0.8$, where $\rho_s = \sqrt{m_i T_e}/(eB)$ is the local ion Larmor radius at sound speed. Wavenumbers are about $15\%$ lower in UNFAV, owing to imperfectly matched flux surfaces (\refig{eq_fav_unfav_reference}).

\subsection{Edge $E_r$ and kinetic profiles}\label{sec:tcv_fav_unfav_profiles}

\begin{figure}[htbp]	
	\centering
		\includegraphics[width=\linewidth]{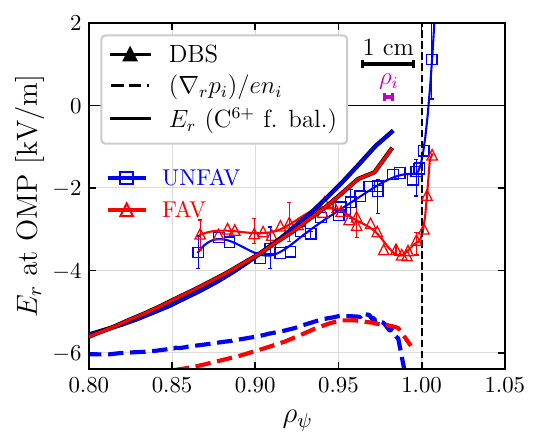}
	  \caption{$E_r = v_\perp B$ profiles inferred from DBS, main ion diamagnetic contribution $(\nabla_r p_i)/(e n_i)$, and $E_r$ derived from the C$^{6+}$ radial force balance. $E_r$ values are mapped to the outboard midplane (OMP).}\label{fig:fav_unfav_vperp_with_vdia}
\end{figure}

\begin{figure}[htbp]
	\centering
		\includegraphics[width=0.9\linewidth]{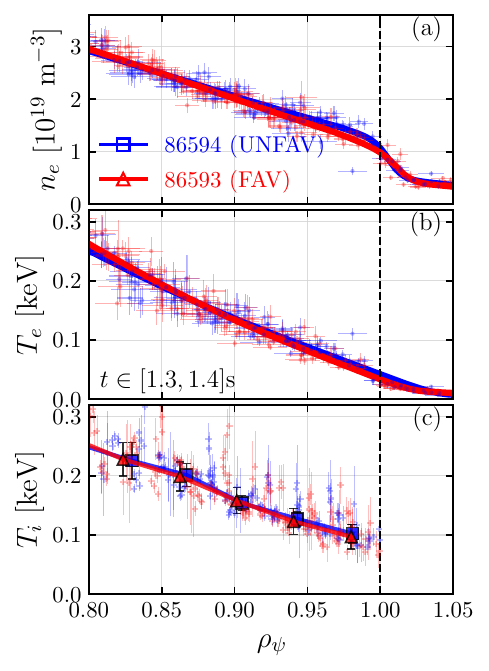}
	  \caption{(a) Electron density and (b) electron temperature from TS, and (c) carbon temperature measured on the LFS by CXRS. Profile fits to the $n_e$ and $T_e$ data are overlaid to guide the eye, while for $T_i$, large markers show the binned weighted average and standard deviation.}\label{fig:fav_unfav_edge_kinetic_profiles}
\end{figure}

\refig{fav_unfav_vperp_with_vdia} shows the DBS data for $E_r = v_\perp B$ mapped to the outboard midplane (OMP),
alongside the $E_r$ profile obtained from the \textit{impurity} (C$^{6+}$) radial force balance, and the \textit{main} ion (deuterium) diamagnetic contribution to $E_r$, $(\nabla_r p_i)/(n_i e)$.
The radial force balance estimation of $E_r$ (C$^{6+}$) relies on a neoclassical estimate of the C$^{6+}$ poloidal velocity using the NEO code\,\cite{Belli_2008}, as described in more details in Ref.\,\cite{Rienaecker_2025}. The diamagnetic term is approximated using carbon temperature as a proxy for deuterium temperature and assuming radially constant main ion dilution, $n_i(r) \propto n_e(r)$, such that $(\nabla_r p_i)/(e n_i) \approx (\nabla_r n_e T_i)/(e n_e)$.
Edge kinetic profiles of electron density ($n_e$), electron and ion temperatures ($T_e$ and $T_i$) are displayed in \refig{fav_unfav_edge_kinetic_profiles}\,(a-c).

In the FAV configuration, the DBS-inferred $E_r$ profile exhibits a pronounced well with a minimum near $\rho_\psi = 0.98$, whereas this well is absent or significantly less pronounced in the UNFAV case. This observation aligns with the trend reported on WEST\,\cite{Vermare_2022} and AUG\,\cite{Schirmer_2006, Plank_2023}.
Meanwhile, density and temperature profiles (\refig{fav_unfav_edge_kinetic_profiles}) are nearly identical, including in the core (not shown).

Examining $E_r$ (C$^{6+}$) in \refig{fav_unfav_vperp_with_vdia}, it shows good agreement with the DBS data up to $\rho_\psi \approx 0.95$. Towards the very edge, unlike the DBS measurements, $E_r$ (C$^{6+}$) does not display a well in either configuration. This suggests that, at the very edge, the radial force balance reconstruction of $E_r$ may currently be inaccurate (as noted previously in \cite{Rienaecker_2025})---probably a result of CXRS diagnostic limitations and/or using a neoclassical estimate for the C$^{6+}$ poloidal rotation.
Next, we consider the \textit{main} ion $(\nabla_r p_i)/(e n_i)$ profiles (not to be confused with the diamagnetic term in the C$^{6+}$ force balance, which is generally much weaker). Consistent with the matched density and temperature profiles, $(\nabla_r p_i)/(e n_i)$ is similar for FAV and UNFAV, up to $\rho_\psi \approx 0.98$.\footnote{Closer to the separatrix, $(\nabla n_e)/n_e$ tends to diverge. For that reason, and considering experimental uncertainties, $(\nabla_r p_i)/(e n_i)$ values for $\rho_\psi \gtrsim 0.98$ should be taken with caution here.} Thus, the TCV results reinforce the findings from AUG\,\cite{Plank_2023}, indicating that the edge $E_r$ well discrepancy between FAV and UNFAV cannot be accounted for solely by differences in $(\nabla_r p_i)/(e n_i)$. 
Furthermore, $(\nabla_r p_i)/(e n_i)$ deviates markedly from $E_r$, consistent with AUG results\,\cite{Plank_2023}, underscoring the significance of the Lorentz contribution (from poloidal and/or toroidal rotation) to the main ion radial force balance. This finding extends beyond the present scenario to a range of conditions in TCV L-mode plasmas, including those with substantial auxiliary heating.

\subsection{Toroidal rotation}\label{sec:tcv_fav_unfav_tor_rotation}

\begin{figure}[htbp]
	\centering
	\includegraphics[width=.9\linewidth]{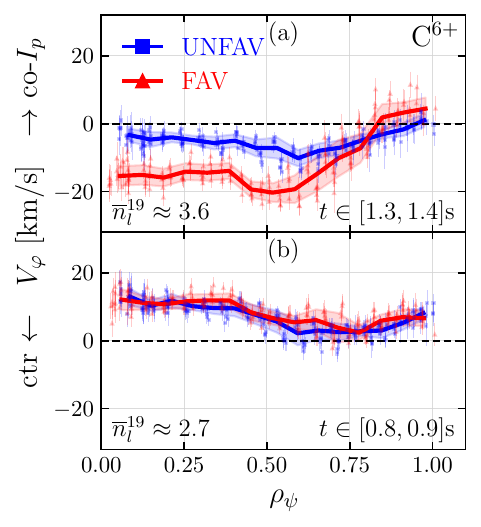}
	\caption{Carbon toroidal rotation from CXRS at the outboard midplane for (a) the reference scenarios and (b) for an earlier, low-density phase of the same discharges. Positive values correspond to co-current rotation. Lines and errorbands represent mean and standard deviation from a binned weighted average over the raw data points.}\label{fig:vtor_IMP_OMP_interm_dens}
\end{figure}

Carbon toroidal rotation ($V_{\varphi}$) profiles, measured at the outboard midplane by CXRS, are shown in \refig{vtor_IMP_OMP_interm_dens} for (a) the time window considered so far and (b) for a different time window.
Examining (a), we note that $V_{\varphi}$ is counter-current in the core, with a stronger magnitude in FAV. Near the edge, however, $V_{\varphi}$ is similar and close to zero in both discharges. In (b), which corresponds to an earlier, low-density phase (line-averaged density, $\overline{n}_l$, between $2.6$ and $\SI{2.8e19}{\per \cubic \m}$) of the same discharges,
the $V_{\varphi}$ profiles are nearly aligned from edge to core, although the edge $E_r$ difference persists.
Other scenarios yield similar results: Depending on plasma parameters, the intrinsic carbon $V_{\varphi}$ in matched FAV and UNFAV discharges sometimes differs significantly in the \textit{core}, even in sign. However, no clear systematic offset is observed, particularly at the edge. CXRS data from the \textit{inboard} midplane (not shown) do not extend to the very high-field-side edge, but are otherwise qualitatively similar to the outboard midplane.
A comparison of carbon poloidal rotation is not provided, as no reliable data are available for the discharges considered in this study.

\subsection{Density fluctuation levels}\label{sec:tcv_fav_unfav_comparison_fluct}

\begin{figure}[htbp]
	\centering
		\begin{subfigure}[b]{0.8\linewidth}
            \includegraphics[width=\linewidth]{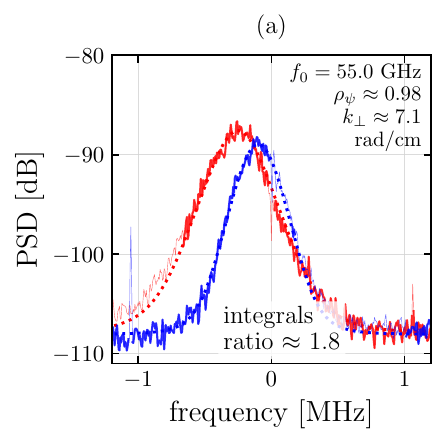}
		\end{subfigure}
		\begin{subfigure}[b]{0.8\linewidth}
			\includegraphics[width=\linewidth]{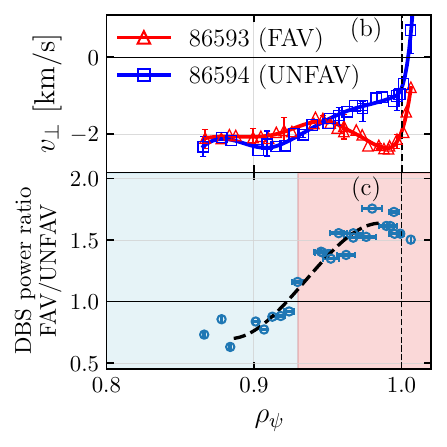}
		\end{subfigure}
	\caption{(a) PSD of the DBS signals compared between FAV and UNFAV at the minimum of the $E_r$ well. Dashed lines indicate fits to the Doppler-shifted peaks. Radial profiles of (b) $v_\perp$ (for reference) and of (c) the ratio of the integral DBS power computed from the fits. At the edge, the higher DBS power in FAV suggests a stronger density fluctuation level $|\tilde{n}(k_\perp)|$.}\label{fig:example_spectra_fav_unfav}
\end{figure}

The intensity of the DBS signal carries localized information on the amplitude of the scattering density fluctuations, $\tilde{n}(k_\perp)$\,\cite{Hennequin_2006}.
We compare the power spectral density (PSD) of the complex DBS signal, whose integrated power serves as a qualitative proxy for $\tilde{n}(k_\perp)$.
Because the DBS signals are not calibrated across different probing frequencies, the radial variation of $\tilde{n}(k_\perp)$ cannot be examined in this analysis. To enable meaningful comparisons of $\tilde{n}(k_\perp)$ between discharges, the DBS probing geometry, radial location ($\rho_\psi$), scattering wavenumber ($k_\perp$), and probing frequency ($f_0$) are held constant. For example, \refig{example_spectra_fav_unfav}(a) presents such a comparison at $\rho_\psi \approx 0.98$.

In addition to the higher Doppler shift observed in FAV—reflecting the deeper $E_r$ well—the Doppler peak is also broader and somewhat higher than in UNFAV. Consequently, the backscattered power, estimated by integrating over the Doppler peak, is higher by a factor of 1.8.
This power ratio, calculated across various probing frequencies, is shown as a function of radial location in \refig{example_spectra_fav_unfav}\,(c). The region where the ratio exceeds one—indicating higher DBS power in FAV—aligns radially with the $E_r$ discrepancy.
This suggests a higher level of edge density fluctuations in FAV within the probed poloidal volume and within the turbulence wavenumber range $k_\perp \rho_s \sim 0.5$--$0.8$.
As previously noted, $k_\perp$ is approximately 15\% lower in UNFAV. Since turbulence power within this $k_\perp$ range decreases with increasing $k_\perp$\cite{Vermare_2011}, the slight mismatch in $k_\perp$ is unlikely to account for the observed difference in DBS power between FAV and UNFAV.
The trend of higher backscattered DBS power in FAV persists qualitatively over various densities, plasma currents, as well as in a comparable, upper single-null (USN) geometry.
For certain pairs\footnote{An example is the high-density, Ohmic LSN discharges at $\SI{250}{kA}$ represented in \refig{tcv_density_scan_LSN}\,(c).}, complementary edge fluctuation data from the midplane gas-puff imaging (GPI\,\cite{Offeddu_2022}) diagnostic are available. Even in the case of significantly higher (up to a factor 2) DBS power in FAV, the fluctuation levels of GPI brightness (not shown), measured at the outboard midplane between $\rho_\psi = 0.95$--$1.025$, do not indicate any notable difference between FAV and UNFAV.

\subsection{Radial correlation length}\label{sec:radcorr}

We extend the comparison of density fluctuation properties in FAV and UNFAV by analyzing the radial correlation length ($\ell_c$) estimated via DBS.
Following the analysis procedure described in \,\cite{Panico_2026_inprep}, the cross-correlation function $\mathcal{C}$ is computed between the DBS amplitude signals collected with the two available channels (reference and hopping frequencies)\,\cite{Panico_2026_inprep}. An exponential decay of the cross-correlation maximum is assumed, according to $\mathcal{C}_\mathrm{max} \propto \exp(-\Delta / \ell_c)$, where $\Delta$ is the approximately radial offset between hopping and reference measurement locations. 
The spatial decay of $\mathcal{C}_\mathrm{max}$ thus yields $\ell_c$, which we take as qualitative proxy of the characteristic radial size of the scattering density perturbations (at a given $k_\perp$).
We restrict the analysis to $\Delta < 0$, such that $\rho_\psi^\up{hop} < \rho_\psi^\up{ref}$, and only consider radial offsets of up to $|\Delta| \le \SI{6}{mm}$.\footnote{Correlation measurements over longer distances are susceptible to the effect of long-range events akin to avalanches\,\cite{Panico_2026_inprep}, not examined here.}
DBS-correlation measurements require dedicated diagnostic settings and stationary plasma conditions over several 100s of milliseconds. Therefore, subsequent results are obtained in a separate pair of discharges. They are largely identical to the references, except for a slightly lower density ($\overline{n}_l \approx \SI{3.1e19}{}$ instead of $\SI{3.6e19}{\per \cubic \m}$).

\begin{figure}[htbp]
	\centering
		\begin{subfigure}[b]{0.8\linewidth}
            \includegraphics[width=\linewidth]{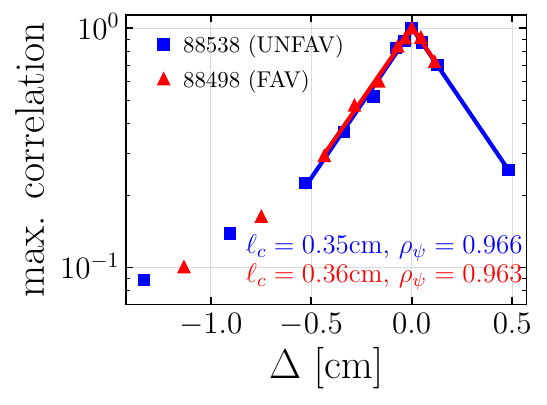}
			\caption{}\label{fig:correl_function_fav_unfav}
		\end{subfigure}
		\begin{subfigure}[b]{0.8\linewidth}
			\includegraphics[width=\linewidth]{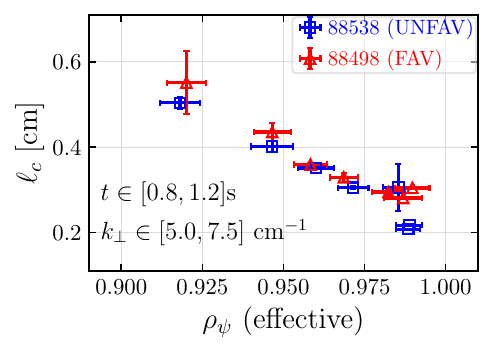}
			\caption{}\label{fig:correl_length_fav_unfav}
		\end{subfigure}
	\caption{(a) Maximum of the cross-correlation function versus radial probing distance between reference and hopping DBS amplitude signals. Annotated values ($\ell_c$ and $\rho_\psi$) are the correlation length obtained from fits to the inner-side decay region ($\Delta \le 0$), and the radial location of the reference signal. (b) Comparison of $\ell_c$ around different (effective) radial locations.}\label{fig:correl_fav_unfav}
\end{figure}

\refig{correl_function_fav_unfav} shows the spatial decay of the correlation maximum around a reference probing radius $\rho_\psi^\up{ref} \approx 0.96$.
It shows very similar decay lengths in UNFAV and FAV ($\ell_c \approx \SI{3.5}{mm}$).
Repeating the analysis over various $\rho_\psi^\up{ref}$ yields radial profiles of $\ell_c$, displayed in \refig{correl_length_fav_unfav} as a function of an effective $\rho_\psi$ as defined in \cite{Panico_2026_inprep}; $\ell_c$ decreases towards the separatrix while remaining similar in the two drift cases. Two additional, comparable discharge pairs (not shown) reveal mild differences in $\ell_c$ (locally up to \SI{2}{mm}) between FAV and UNFAV. However, the available measurements do not indicate a clear, systematic difference at the edge. That said, a rigorous comparison of $\ell_c$ is complicated; Results may be sensitive to slight mismatches in $k_\perp$ or edge density profiles. In addition, the effects of nonlinear scattering\,\cite{Gusakov_2002} at the very edge, or eddy tilting\,\cite{Pinzon_2019a,Pinzon_2019b} may introduce additional, drift-dependent biases.

\section{Plasma current scan}\label{sec:ip_scan}

Next, we investigate how the FAV/UNFAV difference in $E_r$ depends on plasma conditions, beginning with the plasma current magnitude $|I_p|$.
Motivated by experiments on WEST\,\cite{Vermare_2022} demonstrating a marked sensitivity of the edge $E_r$ profile to $|I_p|$ in UNFAV, we perform a corresponding study on TCV. The shot-by-shot scan covers $|I_p|$ values of $[150, 210, 250]$ kA, corresponding to an edge safety factor $q\low{95} \approx 5.0$--$3.0$, comparable to the WEST experiments\,\cite{Vermare_2022}. The reference FAV and UNFAV discharges from \refsec{tcv_fav_unfav} are used as the intermediate $|I_p|$ case. 
In each discharge, a linear density ramp is performed. By selecting appropriate time intervals, the line-averaged density ($\overline{n}_l$) is maintained across the $I_p$ scan to better isolate the effect of $I_p$.

\begin{figure*}[htbp]
	\centering
	\begin{subfigure}[b]{0.8\linewidth}
		\includegraphics[width=\linewidth]{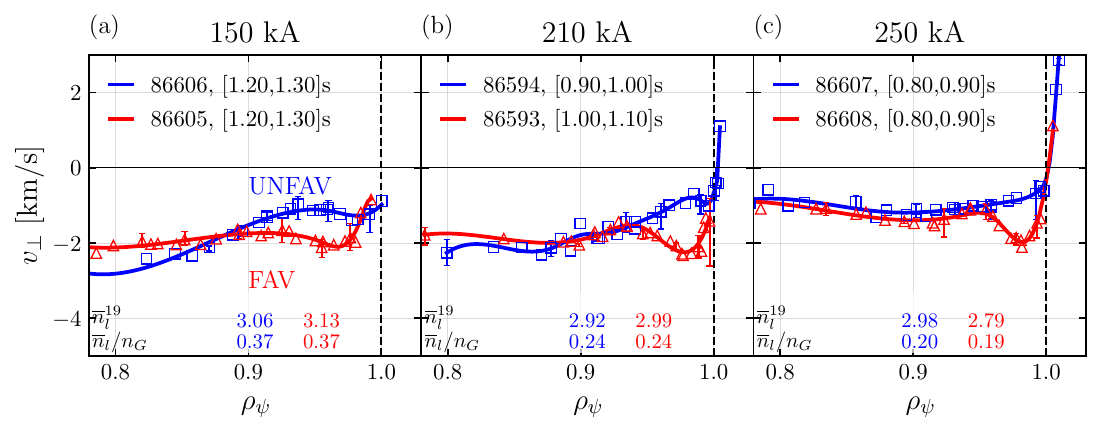}
	\end{subfigure}
	\caption{Perpendicular velocity from DBS in FAV and UNFAV configurations at various $|I_p|$, with fixed line-averaged density ($\overline{n}_l^{19}$; values in [10$^{19}$\,m$^{-3}$] and corresponding Greenwald fractions $\overline{n}_l / n_G$ shown below). No clear trend is observed at the edge.}\label{fig:tcv_ip_scan_LSN}
\end{figure*}
The edge $v_\perp$ profiles at $\overline{n}_l \approx \SI{3e19}{\per \cubic \m}$ are compared in \refig{tcv_ip_scan_LSN}. These profiles show no significant sensitivity to $|I_p|$ in either FAV or UNFAV configuration. Only subtle changes are observed: as $|I_p|$ increases, the well in FAV becomes narrower, but its depth remains largely unchanged.
In UNFAV, increasing $|I_p|$ further flattens the shallow $E_r$ well, which disappears entirely at the highest current. Towards the core ($\rho_\psi \lesssim 0.9$), $E_r$ increases with $|I_p|$ in both drift directions, correlating with a co-current shift in carbon toroidal rotation (not shown).
Repeating the analysis at higher density ($\overline{n}_l \approx \SI{4e19}{\per \cubic \m}$) yields similar qualitative results. Likewise, an $I_p$ scan performed with a different (USN) geometry leads to the same conclusion\,\cite{Rienaecker_2026_phdthesis}, including a recent comparison at even higher $|I_p|=\SI{300}{kA}$.
Thus, in contrast to WEST, $|I_p|$ has little effect on the $E_r$ well in TCV.

\section{Density scan}\label{sec:density_scan}

\begin{figure*}[htbp]
	\centering
    \begin{subfigure}[b]{0.8\linewidth}
        \includegraphics[width=\linewidth]{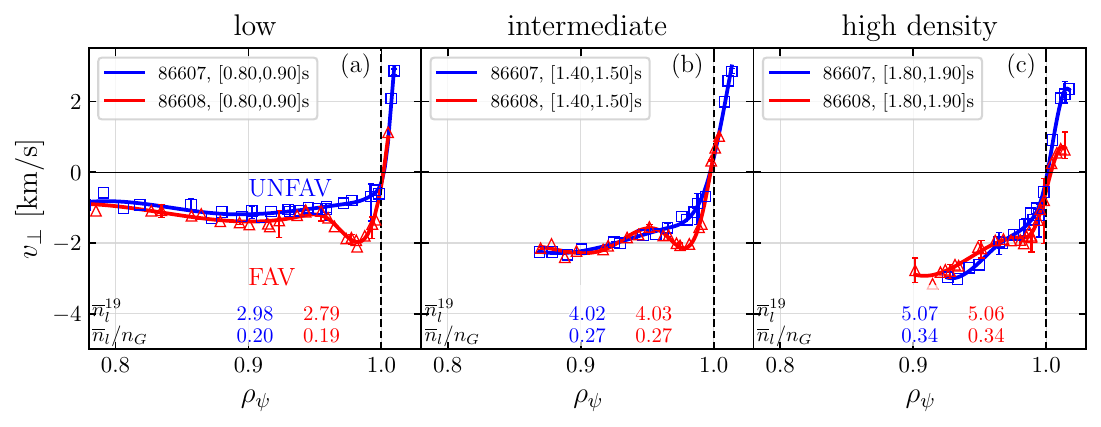}
    \end{subfigure}
    \caption{Perpendicular velocity from DBS compared at various densities, for a LSN scenario at $\SI{250}{kA}$ (same discharges as \refig{tcv_ip_scan_LSN}\,c). Increasing density reduces the FAV/UNFAV difference in edge $E_r$.}\label{fig:tcv_density_scan_LSN}
\end{figure*}

\refig{tcv_density_scan_LSN} compares the $E_r$ profiles from matched FAV and UNFAV Ohmic discharges at various line-averaged densities ($\overline{n}_l$) ranging from 3 to \SI{5e19}{\per \cubic \m}.
Their value ($\overline{n}_l^{19}$) in [$10^{19}$\SI{}{\per \cubic \m}] and the Greenwald fraction ($\overline{n}_l / n_G$) are indicated at the bottom of the panels.
As density increases, the $E_r$ well in the FAV discharge becomes less distinct. In contrast, the UNFAV case, which lacks a well at all densities, shows a pronounced decrease in edge $E_r$ with increasing density, approaching the FAV level. Consequently, at high density, the FAV/UNFAV difference in $E_r$ inside the separatrix becomes small.

\begin{figure}[htbp]
	\centering
    \includegraphics[width=\linewidth]{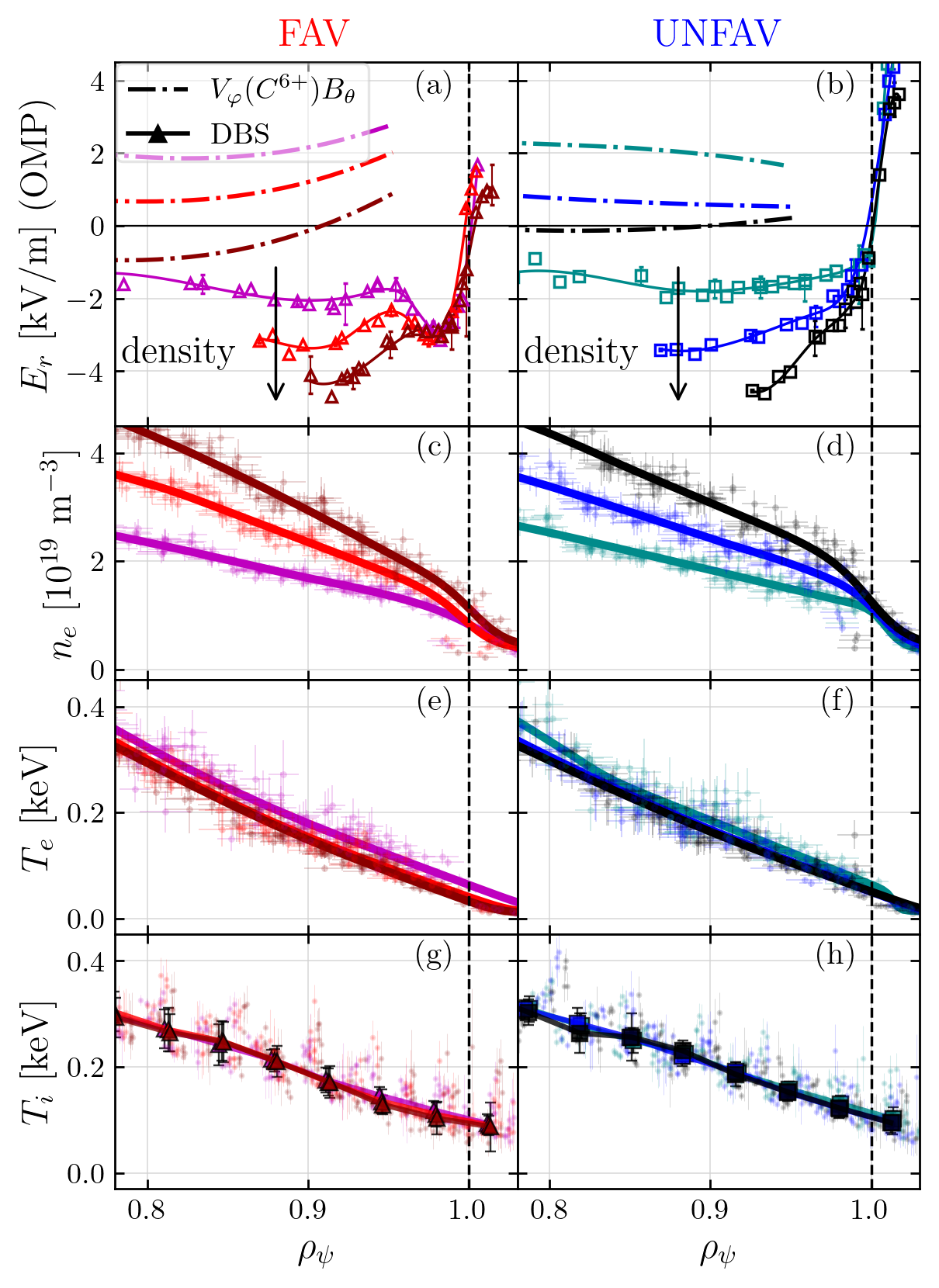}
	\caption{(a-b) DBS data from \refig{tcv_density_scan_LSN} shown as $E_r = B v_\perp$ at the outboard midplane, separated into FAV and UNFAV (left/right). Overlaid is the carbon radial force balance toroidal rotation term, $V_\varphi (C^{6+}) B_\theta$, which decreases with density similarly to the DBS-inferred $E_r$. (c-h) Density, electron temperature, and ion temperature profiles.}\label{fig:LSN_250kA_density_dep}
\end{figure}
For a closer look at this trend, \refig{LSN_250kA_density_dep}(a,b) presents the variation of $E_r$ with density, separated into FAV/UNFAV (left/right panels), together with the density and temperature profiles in (c--h). 
With increasing density, the $E_r$ level decreases in both drift configurations. In the UNFAV case, this decrease extends from the outer core nearly to the separatrix. In contrast, in the FAV case, $E_r$ remains nearly constant near the well minimum. As a result, the shear layer left to the $E_r$ well weakens—and nearly reverses—in FAV, rendering the well less distinct at higher densities.
Similar qualitative results are observed in corresponding USN discharges.
Note that the edge $T_e$ and $T_i$ profiles exhibit only minor or negligible changes across the density scans (\refig{LSN_250kA_density_dep}\,e-h). This underpins the direct (isolated) impact of density on $E_r$ and carbon rotation (discussed below).

The decrease in $E_r$ is accompanied by a systematic counter-current shift in the carbon toroidal velocity ($V_\varphi$), extending from the core to the edge.
The $V_\varphi B_\theta$ contribution to $E_r$ in the impurity radial force balance is shown as dashed lines in \refig{LSN_250kA_density_dep}\,(a,b). 
The magnitude of the $V_\varphi B_\theta$ variation with density can account for the change in $v_\perp$.

\section{Auxiliary heating scan}\label{sec:heating_scan}

Edge $E_r$ and confinement behaviors in FAV and UNFAV configurations are contrasted as they approach the L-H transition. To that end, matched heating scans are performed using either neutral beam injection (NBI) or electron cyclotron resonance heating (ECRH).

\begin{figure*}[htbp]
    \centering
    \begin{subfigure}[b]{0.15\linewidth}
		\includegraphics[width=\linewidth]{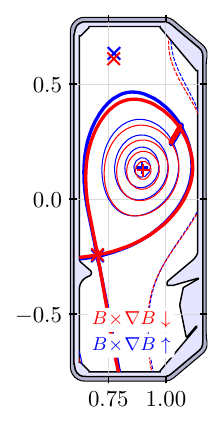}
        \caption{}\label{fig:eq_nbi_lsn_fav_unfav}
	\end{subfigure}
    \centering
    \begin{subfigure}[b]{0.8\linewidth}
        \includegraphics[width=\linewidth]{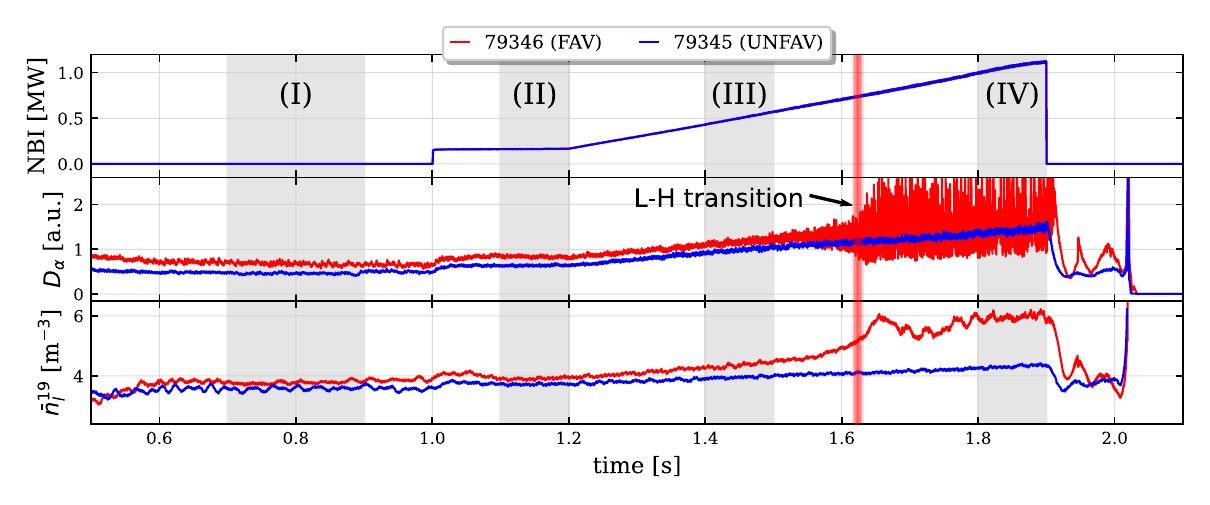}
        \caption{}\label{fig:scopes_nbi_lsn_fav_unfav150kA_annotated}
	\end{subfigure}
    \caption{(a) Equilibria of the NBI-heated FAV and UNFAV discharge pair. (b) From top to bottom: Time traces of NBI power, vertically detected D$_\alpha$ line signal indicating the transition to an ELMy H-mode regime, and line-averaged density. The shaded areas indicate the L-mode time windows investigated. The FAV discharge undergoes an L-H transition around \SI{1.63}{s}.}\label{fig:NBI_ramps_signals_150kA}
\end{figure*}

In the NBI case, a pair of LSN equilibria (\refig{eq_nbi_lsn_fav_unfav}) is used that is similar to the Ohmic references of \refsec{tcv_fav_unfav}, but with $|I_p| = \SI{150}{kA}$ and vertically offset by about $\SI{+13}{cm}$\footnote{The upward-shifted equilibrium was motivated by DBS probing constraints at that time.}.
The heating power is delivered by a co-current neutral beam (NBI-1\,\cite{Karpushov_2023}) and ramped up as shown in \refig{scopes_nbi_lsn_fav_unfav150kA_annotated}. The same (co-current) beam is used in the two drift cases, keeping the sign of $I_p$ fixed and flipping only $B_0$.\footnote{The resulting reversal of magnetic helicity was shown to have no sizeable effect on $E_r$\,\cite{Rienaecker_2025} in a similar, Ohmic scenario.}
The L-mode phases examined along the ramp include: (I) an initial Ohmic phase, (II) an intermediate power step, and (III) a higher-power phase. In the FAV discharge, phase (III) is shortly followed by an L-H transition around \SI{1.6}{s} at an injected power of $P\low{NBI}=700$--$\SI{750}{kW}$. The UNFAV discharge remains in L-mode throughout the ramp.
\begin{figure*}[htbp]
    \centering
    \includegraphics[width=.9\linewidth]{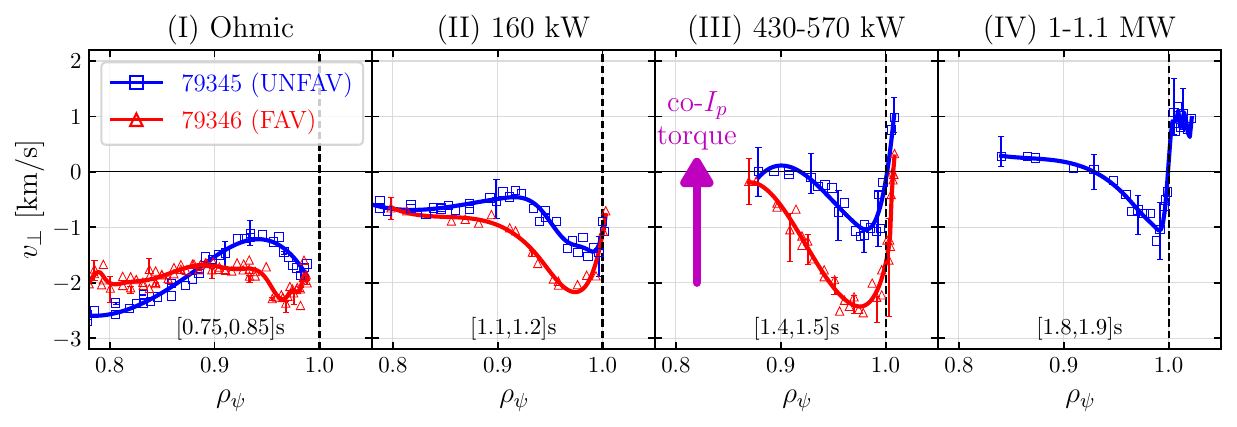}
    \caption{Perpendicular velocity from DBS for the phases displayed in \refig{NBI_ramps_signals_150kA}.}\label{fig:150kA_NBI_comparison}
\end{figure*}
The corresponding $v_\perp$ measurements are displayed in \refig{150kA_NBI_comparison}.
In both drift configurations, the outer core $E_r$ level ($\rho_\psi \lesssim 0.9$) increases with power as a result of the co-$I_p$ torque---an effect consistently observed when using co-$I_p$ NBI\,\cite{Rienaecker_2025}.
Focusing on the \textit{edge} $E_r$ response to the NBI, it is somewhat different between FAV and UNFAV (\refig{150kA_NBI_comparison}): In FAV, the initial well depth is maintained along the ramp (against the increasing co-$I_p$ torque). By contrast, the well becomes shallower in the UNFAV case. 
Even around maximum beam power, $P\low{NBI}=1$--$\SI{1.1}{MW}$, the $E_r$ well remains shallow in UNFAV (\refig{150kA_NBI_comparison}-IV).
An analogous NBI scan replacing the LSN shapes by their USN counterpart yields qualitatively similar results\,\cite{Rienaecker_2026_phdthesis}. 

\begin{figure*}[htbp]
    \centering
    \begin{subfigure}[b]{0.15\linewidth}
		\includegraphics[width=\linewidth]{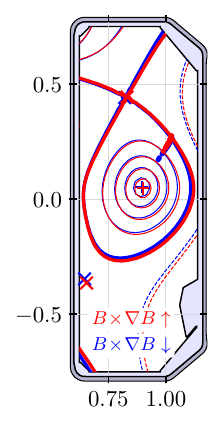}
        \caption{}\label{fig:eq_ecrh_fav_unfav}
	\end{subfigure}
    \centering
    \begin{subfigure}[b]{0.84\linewidth}
        \includegraphics[width=\linewidth]{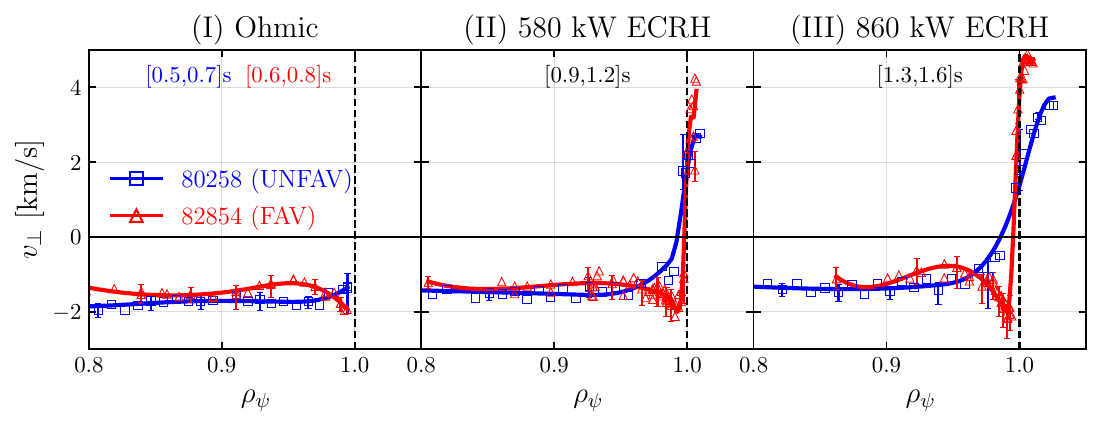}
	    \caption{}\label{fig:vperp_ecrh_fav_unfav}
    \end{subfigure}
    \caption{(a) USN plasma geometries used for the ECRH power scan. (b) Perpendicular velocity from DBS for three consecutive power steps in L-mode.}\label{fig:ecrh_fav_unfav}
\end{figure*}

Next, discharges heated by second-harmonic (X2) ECRH are considered. Here, the plasma scenario is different from the NBI case, including in geometry (\refig{eq_ecrh_fav_unfav}). The main parameters are: $B_0= \colpslashm \SI{1.43}{T}$, $I_p= \colpslashm \SI{170}{kA}$, $q_{95} \approx 4$, $\overline{n}_l = 2.3$--$\SI{3.3e19}{\per \cubic \m}$.
The power of one gyrotron is increased in steps, resulting in central power deposition and full absorption according to raytracing calculations using the TORAY-GA code\,\cite{Kritz_1982}.
Again, three successive L-mode phases are examined: (I) an initial Ohmic phase, (II) a first ECRH step at $P\low{ECRH}=580$, and (III) a second one at $\SI{860}{kW}$. 
The $v_\perp$ profiles are displayed in \refig{vperp_ecrh_fav_unfav}.
As the ECRH power increases, the $E_r$ well depth is maintained, or slightly enhanced, in the FAV discharge, whereas in the UNFAV case $E_r$ just inside the separatrix tends to increase---qualitatively consistent with the NBI case.
In addition, the outer-side shear (to the right of the $E_r$ well) is further intensified in FAV by a sharper jump of $E_r$ across the separatrix.

\begin{figure*}[htbp]
    \centering
    \begin{subfigure}[b]{0.495\linewidth}
		\includegraphics[width=\linewidth]{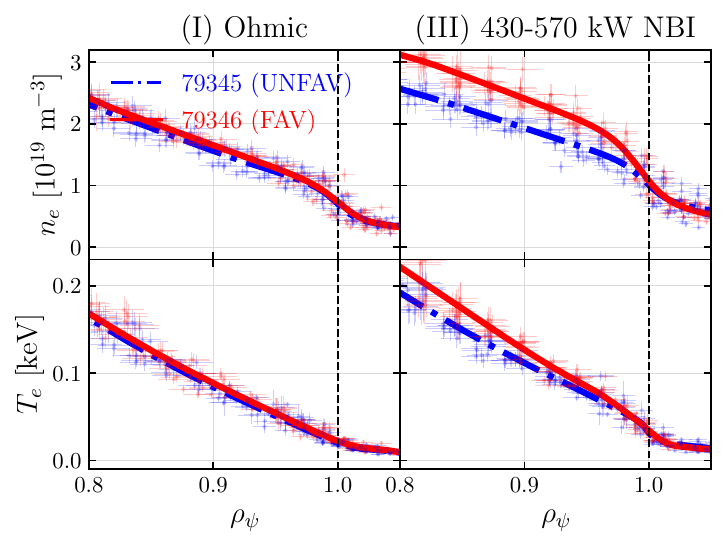}
        \caption{}\label{fig:kinprofs_nbi_fav_unfav}
	\end{subfigure}
    \centering
    \begin{subfigure}[b]{0.495\linewidth}
        \includegraphics[width=\linewidth]{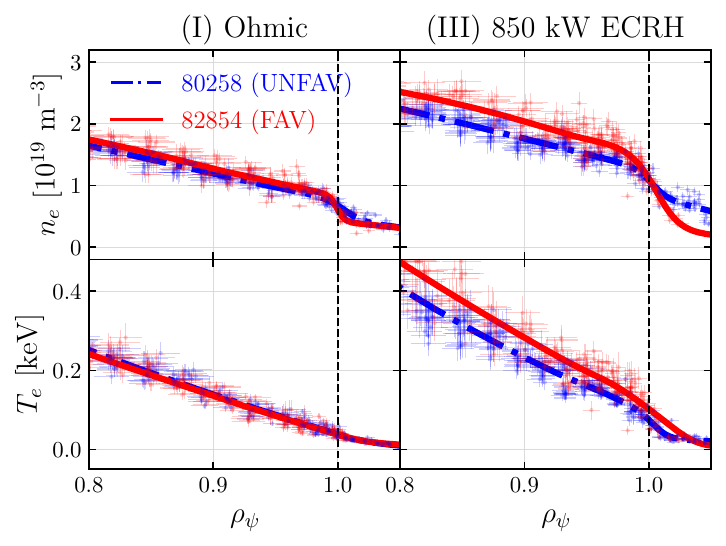}
	    \caption{}\label{fig:kinprofs_ecrh_fav_unfav}
    \end{subfigure}
    \caption{Electron density and temperature profiles (upper and lower panels) corresponding to (I) the Ohmic and (III) the high-power L-mode phases of the (a) NBI and (b) ECRH scans.}\label{fig:kinprofs_fav_unfav}
\end{figure*}

\refig{kinprofs_fav_unfav} compares the edge density and electron temperature profiles\footnote{Good-quality edge $T_i$ data are not available here.} during (I) the low power and (III) the high power phases, from the NBI and ECRH experiments, respectively (a,b).
NBI and ECRH scans again reveal a consistent trend: Initially, under Ohmic heating, $n_e$ and $T_e$ are closely matched between FAV and UNFAV, in agreement with the reference discharges (\refig{fav_unfav_edge_kinetic_profiles}). 
With the additional NBI or ECRH, however, the buildup of the edge $n_e$ and $T_e$ profiles and their gradients is accentuated in the FAV relative to the UNFAV drift case. The increasing edge pressure difference indicates that transport is locally reduced in FAV---as reflected by improved global confinement---compared to UNFAV.

Combining the NBI and ECRH experiments, which cover complementary plasma conditions, we conclude: The L-mode $\ErxB$ shear and confinement at the edge are more readily enhanced by auxiliary heating in the FAV than in the UNFAV drift configuration. 

\begin{figure}[htbp]
	\centering
    \includegraphics[width=0.8\linewidth]{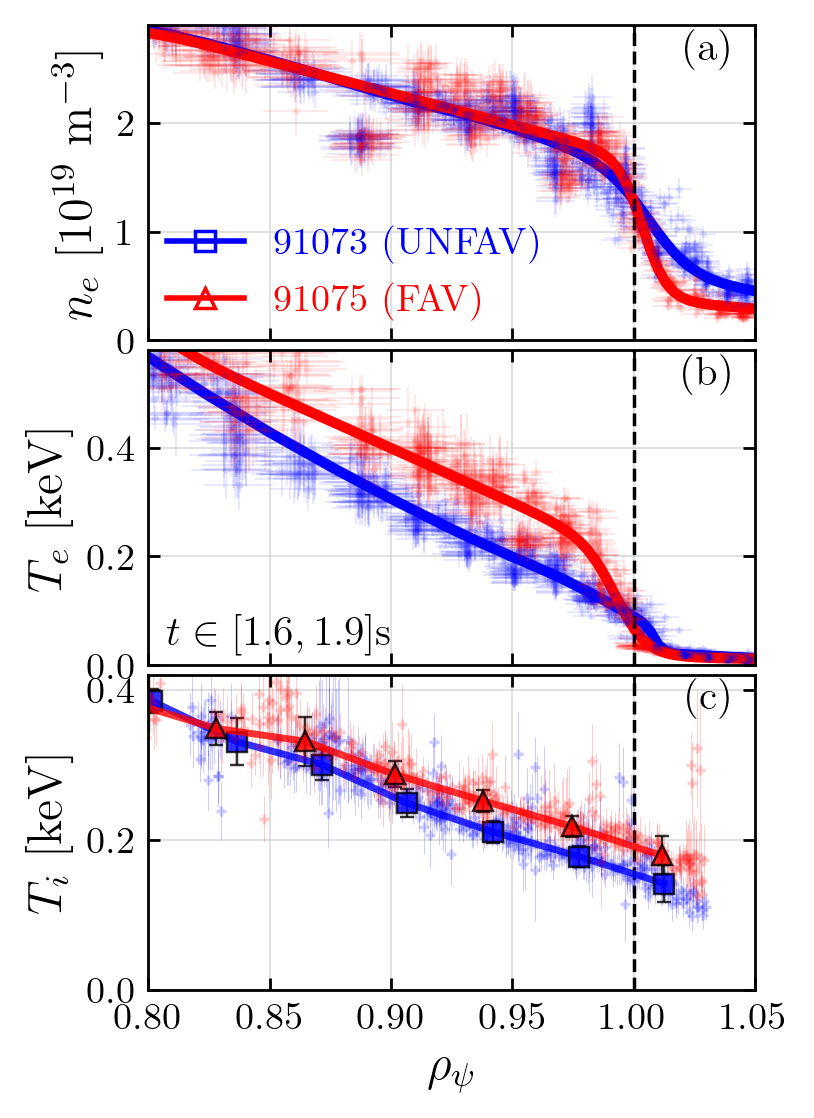}
	  \caption{Edge kinetic profiles (same representation as in \refig{fav_unfav_edge_kinetic_profiles}) for a pair of auxiliary-heated FAV/UNFAV discharges ($P\low{ECRH}=\SI{900}{kW}$) with matched density profiles.}\label{fig:USN_ecrh_edge_kinetic_profiles}
    \end{figure}

\begin{figure}[htbp]
        \centering
        \begin{subfigure}[b]{0.8\linewidth}
            \includegraphics[width=\linewidth]{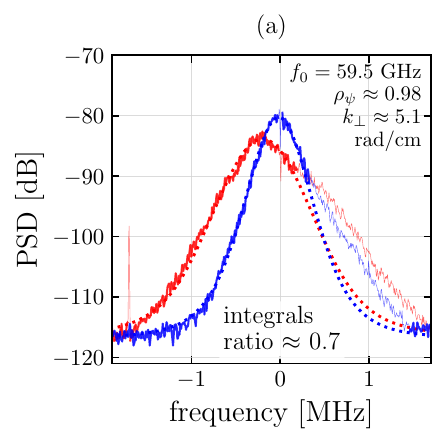}
        \end{subfigure}
        \centering
    \begin{subfigure}[b]{0.8\linewidth}
        \includegraphics[width=\linewidth]{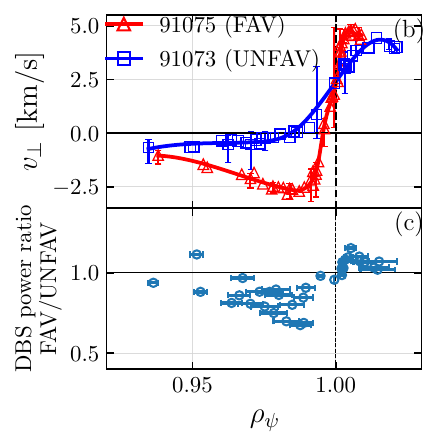}
    \end{subfigure}
    \caption{Same representation as in \refig{example_spectra_fav_unfav}, but for an EC-heated scenario. Unlike the Ohmic case, the DBS power at the edge is lower in FAV.}\label{fig:psd_fav_unfav_ecrh}
\end{figure}

The NBI and ECRH scans considered do not permit a meaningful comparison of DBS fluctuation levels in FAV and UNFAV\footnote{The DBS power is not comparable here due to different DBS settings in FAV and UNFAV, in addition to differing density profiles at same auxiliary heating power.}. To investigate fluctuation levels under auxiliary heating, we resort to a different scenario: USN plasmas with $|I_p| = \SI{250}{kA}$ exposed to \SI{900}{kW} of ECRH power, which is calculated to be fully absorbed in both drift cases. 
The edge $n_e$ profiles are matched (\refig{USN_ecrh_edge_kinetic_profiles}\,a), which is necessary for comparing DBS power, while $T_e$ and $T_i$ are higher in FAV (b,c). The edge $E_r$ profiles (\refig{psd_fav_unfav_ecrh}\,b) differ more substantially than with reduced or purely Ohmic heating, reinforcing the previously noted trends.
\refig{psd_fav_unfav_ecrh}\,(a) compares the DBS frequency spectra at $\rho_\psi \approx 0.98$. 
The Doppler peak is lower in FAV, which is qualitatively opposite to the Ohmic LSN reference case (\refig{example_spectra_fav_unfav}) and clearly contrasts, too, with an Ohmic counterpart to the present scenario (not shown).
The reduced DBS power in FAV, corresponding to a ratio $<1$ in \refig{psd_fav_unfav_ecrh}\,(c), is observed inside the separatrix, most notably at the FAV $E_r$ well minimum.

\section{Discussion}\label{sec:discussion}

The FAV/UNFAV difference in edge $E_r$ observed in TCV L-mode plasmas qualitatively agrees with previous studies on WEST\,\cite{Vermare_2022} and AUG\,\cite{Plank_2023}, underscoring its universality. Notably, for Ohmic discharges a clear impact of drift configuration on $E_r$ can be observed even under well-matched edge density and temperature profiles (\refig{fav_unfav_edge_kinetic_profiles}).
Consistent with findings from AUG\,\cite{Plank_2023}, our results suggest that the edge $E_r$ difference does not merely result from the main ion $(\nabla p_i)/e n_i$ contribution.
Nor is it reflected by a systematic dependence of the edge carbon toroidal rotation on drift configuration (\refig{vtor_IMP_OMP_interm_dens}). This last observation contrasts with Alcator C-Mod\,\cite{LaBombard_2004,LaBombard_2005, LaBombard_2008} and aligns with AUG results\,\cite{Plank_2023}. Thus, neither TCV nor AUG experiments support the interpretation proposed by LaBombard \textit{et al.}\,\cite{LaBombard_2004}, which attributes the drift-dependence of $\PLH$ to toroidal rotation boundary conditions set by transport-driven parallel SOL flows.
Taken together, these observations are compatible with non-neoclassical main ion poloidal rotation being the primary contributor to the edge $E_r$ sensitivity to $\BxgradB$, as suggested by recent \texttt{GENE-X} simulations\,\cite{Frei_2026}.      

Unlike in WEST, the TCV edge $E_r$ structure is relatively insensitive to $I_p$ (\refig{tcv_ip_scan_LSN}).
Likewise, a restricted $I_p$ scan in UNFAV drift configuration on AUG\,\cite{Rienaecker_2026_phdthesis} reveals only minor variations in $E_r$ well depth.
This discrepancy motivated additional, very recent experiments on TCV to determine whether a WEST-like trend emerges at lower collisionality. So far, however, further increasing $|I_p|$ to $\SI{300}{kA}$ or applying ECRH heating to a $\SI{250}{kA}$ scenario does not qualitatively alter the outcome compared to what is reported here. Thus, it remains unclear which ingredient governs the formation of a deep $E_r$ well at higher $I_p$ in WEST, or why this effect appears to be absent in TCV and AUG.

Increasing density reduces the edge $E_r$ disparity between FAV and UNFAV in TCV (\refig{tcv_density_scan_LSN}), a trend that, to our knowledge, has not been previously reported and thus warrants cross-device confirmation.
In both drift configurations, the macroscopic $E_r$ level decreases with increasing density. The concomitant counter-current shift in toroidal rotation, which is consistent with previous observations in TCV\,\cite[Fig.\,15]{Duval_2008} and AUG\,\cite[Fig.\,19]{Plank_2023}, indicates that the density dependence of $E_r$ may be partly driven by intrinsic toroidal rotation changes. However, caution is required, as impurity rotation may not reflect main ion behavior\,\cite{Kim_1994,Haskey_2018}.
Various mechanisms may directly or indirectly affect the edge $E_r$ and intrinsic rotation, including plasma-neutral interaction\,\cite{Itoh_1989,MonierGarbet_1997,Ippolito_2002,Versloot_2011,Omotani_2016}, collisionality-driven changes in Pfirsch-Schlüter transport\,\cite{Aydemir_2009}, ion-orbit losses\,\cite{Brzozowski_2019}, or turbulence regime\,\cite{Krutkin_2026_inprep}. 
Which of these effects (if any) may explain the reduced FAV/UNFAV difference in $E_r$ at high density, is an open question left for future studies.

It is noteworthy that edge density and temperature profiles, as well as radial correlation lengths (\refig{correl_fav_unfav}), are similar in Ohmic FAV and UNFAV plasmas, despite clear $E_r$ differences.
A plausible explanation could be that, under Ohmic heating, the $E_r$ shear difference between FAV and UNFAV may not significantly affect turbulent transport, nor the radial structure of density perturbations (at least within the probed $k_\perp$ range).
Surprisingly, the concurrent higher DBS power in FAV compared to UNFAV Ohmic plasmas (\refig{example_spectra_fav_unfav}) indicates stronger density fluctuations at larger flow shear.
Meanwile, as noted previously, the FAV/UNFAV discrepancy in DBS power is not reflected by GPI fluctuation levels. 
This may be due to different wavenumber sensitivities of the two diagnostics.\footnote{The GPI setup based on APDs in these experiments has focal spot sizes of $\SI{4}{mm}$, translating to a sensitivity to wavenumbers smaller than $k_\perp \lesssim \SI{1.6}{\per \centi \m}$, compared to $4$--$\SI{10}{\per \centi \m}$ in the case of DBS.}
An analogous analysis of DBS signals in Tore Supra plasmas\,\cite{Hennequin_2010} yields qualitatively similar results\,\cite{Rienaecker_2026_phdthesis}. Likewise, recent Ohmic discharges on WEST\,\cite{Orlacchio_Vermare_2026_priv_comm} exhibit a trend consistent with TCV. 
Measurements on AUG using CECE reveal reduced $T_e$ fluctuation levels in FAV\,\cite{Bielajew_2023b}, associated with a stronger weakly coherent mode (WCM) feature in UNFAV.  
We note that results may vary qualitatively depending on fluctuation quantity (density or temperature) and scale ($k_\perp$), poloidal location on a flux surface, as well as plasma conditions.
Thus, a comprehensive characterization of turbulence using various edge fluctuation diagnostics is necessary to more confidently assess the FAV/UNFAV turbulence behavior and its correlation with $E_r$.

As TCV plasmas approach the L-H transition, the difference in edge $E_r$ between FAV and UNFAV becomes increasingly pronounced (\refig{150kA_NBI_comparison} and \ref{fig:ecrh_fav_unfav}), in parallel with a growing asymmetry in edge pressure and its gradient (\refig{kinprofs_fav_unfav}).
This behavior supports a connection between edge $\ErxB$ shear and L-mode confinement level, as observed also in the context of negative versus positive triangularity shaping\,\cite{Rienaecker_2026,Rienaecker_2026_EPS}. In addition, DBS fluctuation levels are reduced in FAV (\refig{psd_fav_unfav_ecrh})---unlike the trend observed in Ohmic heating (\refig{example_spectra_fav_unfav}). This may indicate that the expected stabilization of turbulence in FAV compared to UNFAV becomes noticeable only (or more clearly) with increased heating. The latter enhances the FAV/UNFAV $E_r$ asymmetry, and may potentially also change turbulence properties in a way that leads to more efficient turbulence saturation by the edge flow shear layer in the FAV drift case. 
Given that transport in L-mode likely influences H-mode access\,\cite{Labit_2025}, our observations are consistent with edge $\ErxB$ behavior playing a role in reducing or increasing $\PLH$ in FAV and UNFAV, though they do not, by themselves, demonstrate such a causality. 
Moreover, the relevance of \textit{mean} $\ErxB$ shear in enabling turbulence suppression at the L-H transition, as opposed to fluctuating or intermittent flow shear, remains unclear in light of contrasting observations across different devices\,\cite{Cavedon_2020,Plank_2022, Silva_2021, Silva_2026_EPS, Estrada_2009, Meyer_2011}.
Although the analysis of $\ErxB$ flow dynamics is beyond the present scope, coherent $v_\perp$ oscillations observed via DBS indicate stronger GAM\,\cite{Winsor_1968,Conway_2021} activity in the UNFAV compared to the FAV drift case\,\cite{Vermare_2026_EPS}.  This tendency appears to correlate more generally with the mean $E_r$ profile, extending beyond the FAV/UNFAV asymmetry, and will be addressed in a dedicated contribution\,\cite{Vermare_GAM_inprep}.

\section{Summary and outlook}\label{sec:summary}

TCV experiments broadly confirm the $\BxgradB$ drift influence on the edge $E_r$ structure, in line with previous WEST and AUG experiments. 
The FAV/UNFAV matched comparisons cover a variety of scenarios, underpinning the robustness of the effect across different machines and plasma conditions.
Density and auxiliary heating modulate the FAV/UNFAV difference in $E_r$, while plasma current shows only a minor impact.
The comprehensive dataset enables detailed comparisons with simulations. Corresponding studies involving the present TCV data, as well as related studies, are presented elsewhere\,\cite{Frei_2026_fav_unfav_in_prep, ElSaifi_inprep, Lambresa_inprep, Krutkin_2026_inprep}.
Such joint analysis is critical for disentangling $E_r$ drives and identifying the origin of the FAV/UNFAV asymmetry.
The availability of edge turbulence simulations also calls for full-wave synthetic DBS diagnostics\,\cite{Krutkin_2019_NF, Lechte_2020b,Hoefler_2025,Orlacchio_2026_EPS}. This, alongside more extensive turbulence characterizations, could lend additional weight to simulation-experiment comparisons.
Future experimental work using the DBS will focus on turbulence and flow \textit{dynamics}---rather than mean characteristics alone.

\appendix 

\ack

This work has been carried out within the framework of the EUROfusion Consortium, partially funded by the European Union via the Euratom Research and Training Programme (Grant Agreement No 101052200 — EUROfusion). 
This work has benefited from a grant managed by the Agence Nationale
de la Recherche (ANR), as part of the program \enquote{Investissements
d'Avenir} under the reference (ANR-18-EURE-0014). 
Views and opinions expressed are however those of the author(s) only and do not  necessarily reflect those of the European Union, the European Commission. Neither the European Union nor the European Commission can be held responsible for them.
This work was supported in part by the Swiss National Science Foundation.

\printbibliography

\end{document}